# Pore pressure study in the Irpinia area (Southern Apennines, Italy)

Vitagliano E.*, Improta L., Pizzino L. and D'Agostino N.

*Istituto Nazionale di Geofisica e Vulcanologia (Rome, Italy)*

*eleonora.vitagliano@ingv.it

**Abstract**
Subsurface pore pressure studies are crucial for understanding the geo-mechanical behaviours of the geological formations and for preventing the failure conditions of the rocks. Although the interplay between pore pressure changes and rock deformation is nowadays widely treated in the literature, the magnitude and the distribution of the fluid pressure regimes at depth in static and dynamic conditions is not completely clear, especially in those areas, such as the fold and thrust belts, characterised by a complex tectonostratigraphic setting. The proposed study deals with the subsurface fluid dynamics of the Irpinia region, located in the Southern Apennines (Italy) and marked by intense tectonic activity and seismicity. Indeed, past earthquakes have played a significant role in shaping the landscape of this region and structuring the geological units. The most recent and notable Italian earthquake occurred in November 1980 (6.9 Mw) and caused significant damage and loss of life. In addition, the Irpinia area is a site of deep gas rising to the surface and exhibits clear correlations between crustal deformation and groundwater circulation. As recently demonstrated by some of the authors, the hydrological forcing causes observable variations in the crustal volumes. The pressure analysis herein proposed has been performed using direct and indirect pressure measurements collected from 13 hydrocarbon exploration wells available in open source. It represents a sort of "reality test" if compared with the hypotheses on fluid pressures and overpressures based on other geophysical methods. In particular, the proposed study provides a detailed description of the methodology used to identify where overpressures develop within the sediments of both autochthonous and allochthonous layers. The study also investigates the relationship between pore pressures, gas occurrences found at well sites, and the possible causes that generate overpressures. Some of the study's significant findings include: i) the carbonate successions of the South-Apennines and Apulian Platforms, which are exposed and buried, respectively, are characterized by predominantly hydrostatic pressure regimes; ii) the shale-rich successions of the Lagonegrese pelagic basin and the Miocene-Pliocene foredeep basin locally demonstrate moderate overpressured gradients; iii) the highest overpressure values are observed in the evaporitic deposits and Pliocene shales.

**1. Introduction**

Understanding the pore pressure, or the fluid pressure within porous rock, is crucial in studying the deformation of the Earth's crust and the mechanical behaviour of geological formations. An increase in pore pressure is known to reduce the effective normal stress, weakening the fault zone and leading the shear strength to a level below the prevailing shear stress (Byerlee J., 1993; Cappa et al., 2009). Pore pressure transients and associated fluid flow increase the frequency and magnitude of earthquakes, as documented by seismicity extraction/injection-induced, such as hydrocarbon exploitation or geological $CO_2$ storage (Segall, 1989; Frohlich et al., 2016; White and Foxall, 2016; Vilarrasa et al., 2019). The recent hydro-shear and hydro-fracturing underground laboratory experiments clearly show that pore pressure variations and fluid flow exert a control over the mechanical and hydraulic properties of rock mass and fault zones, changing the local stress field, and guiding the preparation, initiation, and evolution of rupture mechanisms. Unlike underground experiments and hydrocarbon exploitation activities, which are supported by accurate knowledge of the local structural setting, pressure conditions and physical, mechanical and hydraulic properties of reservoir, the study of natural seismicity suffers from the lack of reliable information on the mechanical and hydraulic properties of the crustal rocks, as well as on the crustal fluids (type, phase, and pressure).

The oil and gas industry has historically collected pore pressure data to achieve mineralization targets while ensuring safe drilling practices (e.g., Zhang, 2011). As a result, the knowledge and terminology on this subject were acquired through well-log analysis and then exported to other disciplines. Nowadays, it is well understood that abnormal pressures, which are pore pressures that deviate from hydrostatic (e.g., Neuzil, 1995), can be caused by various physical processes and exist in different geological environments (e.g., Yassir and Addis, 2001; Swarbrick et al., 2002). Both compressional and extensional geodynamic settings can generate natural mechanisms that lead to overpressures. These mechanisms can be a result of chemical, physical and thermal processes that occur within a rock system. They can also be caused by external forcing of the system, such as sediment/tectonic loading and tectonic stress. The former group includes diagenetic processes, hydrocarbon generation, and pressuring due to pore fluid thermal expansion (Barker, 1972; Rieke and Chilingarian, 1974; Bruce, 1984; Luo and Vasseur, 1996; Lahann, 2002; Guo et al., 2010). The latter group consists of loading processes, associated with porosity reduction, and the pressuring related to the tectonic stress (Hubbert and Rubey, 1959; Rieke and Chilingarian, 1974; Davis et al., 1983; Yassir, 1990; Luo et al., 2007). Moreover, it is widely believed that the distribution and magnitude of overpressures depend on the interplay between pressure formation and dissipation, and that both are related to geological processes (Bredehoeft and Hanshaw, 1968; Neuzil, 1995; Osborne and Swarbrick, 1997; Madon, 2007; O'Connor et al., 2008). Some authors state that the pore fluid is contained in impermeable layers and its abnormal pressure is maintained through time by barriers that prevent fluid flow (e.g., Bradley, 1975; Hunt, 1990; Bradley and Powely, 1994; Ortoleva, 1994). Others assume that abnormal pressures are time-dependent because they are activated by specific and time-confined geological processes (e.g., Bethke, 1986; Bredehoeft et al., 1994; Audet, 1995; Gordon and Flemings, 1998). Finally, concerning the disequilibrium state in a rock system, it can be associated with the chemically- and gravity-driven flows (e.g., Olsen, 1972; Toth, 2009) and also be related to the tectonic environments, wherein a fluid-driven seismicity acts at local or regional scales (e.g., Sibson, 2014) and a transient growth of pore pressures can affect the seismogenic zone (e.g., Chiarabba, 2020).

Several studies suggest that there are high pore pressure zones and fluid movements within the upper crust of the Apennine chain. These overpressures are believed to be the primary cause of recent activation of normal faults that have resulted in large earthquakes such as Colfiorito, L'Aquila, Amatrice-Visso-Norcia, and Matese (M>6). The spatial-temporal evolution of the associated seismic sequences is also attributed to these overpressures. The conceptual model is based on fluid-driven seismicity, where deep $CO_2$-rich fluids from the middle crust upwell locally, and mixed $H_2O$-$CO_2$ fluids are trapped in upper crust reservoirs (such as fractured carbonates). The frictional behavior of the rocks depends on the compressibility of these fluids. This geophysical-based model is supported by indirect data, including the seismic velocity anomaly (Vp/Vs ratio) observed at the base of the seismogenic zone before and after mainshocks (e.g. Chiarabba et al. 2020). Similar mechanisms have been hypothesized for the Irpinia earthquake of 1980 Ms6.9, based on the correlation found between the spatial-temporal patterns of recent background seismicity and seismicity of the small sequences, and on the physical properties of Apulian carbonate rocks inferred from 3D and time-lapse seismic tomography surveys (Amoroso et al., 2014; Improta et al., 2014; Amoroso et al., 2017; De Landro et al., 2022). Recent analyses of background seismicity in the Irpinia region have also shown a correlation between seismicity rates, ground deformation inferred from geodetic data, and the seasonal variation of large karst aquifers (e.g., D'Agostino et al., 2018; Silverii et al., 2019).

However, the interpretation of both large seismicity sequences and background seismicity for the entire Apennines (particularly for the Irpinia region) and the control of crustal fluids on the stress variations remain uncertain. Little is known about the hydraulic properties of sedimentary rocks at seismogenic depths, pore pressure patterns, and fluid flows in the highly heterogeneous sedimentary crust of the Apennine thrust-and-fold belt. In particular, there is uncertainty about where the overpressures are localized in the chain and what the generation processes are. Another issue is the contribution of groundwater circulation through the shallow karst system to the overpressure. Furthermore, it is essential to investigate how the deep arrival of fluids, such as hydrocarbons and carbon dioxide, can affect the pressure regime in the deep carbonate reservoirs. Ultimately, it is critical to understand how pressure is dissipated at the regional scale and through which mechanisms. Our research delved into the pore pressure distribution and magnitude within the geological formations of the Irpinia area up to a depth of 6 km. The primary objective of our work is to clarify some of these aspects to gain a better understanding of the effects of pore pressures and fluid flows in the study area.

## 2. Geology and fluids in the study area

The Irpinia region in the Southern Apennine mountain range developed over the last 20 million years due to the convergence of the European and African tectonic plates along a west-directed subduction zone (Doglioni et al., 1996; Carminati et al., 2004; Vezzani et al., 2010 and reference therein). This zone progressively moves eastward. The current tectonic configuration in the area reflects the extensive tectonic activity that occurred in the Mediterranean region during the Mesozoic era (e.g., Patacca and Scandone, 2007 and reference therein). Southern Italy experienced crustal extension, thinning, and significant subsidence since the upper Triassic period. This created a morphology characterized by pelagic basins surrounded by extensive carbonate platforms. In the late Oligocene period, the area showed an alternation of deep- and shallow-water depositional domains. These domains include the Liguride-Sicilide Basins, the Apennine Platform, the Lagonegro-Molise Basin, and the Apulian Platform from west to east. Starting from the lower Miocene era, sedimentary units deposited in these domains were progressively involved in the tectonic accretion process, resulting in the structuration of the thrust belt (Scrocca, 2010). At the same time, a series of foredeep basins were involved in compressional tectonic processes while wedge-top basins became younger eastwards. This demonstrates the gradual propagation of the contractional front towards the Apulian foreland (Mazzoli et al., 2013). By the middle Pliocene era, the complex Apennine nappe reached and overlaid the Apulia Platform. This heavily deformed the platform through compressional tectonics, partly reactivating pre-existing Mesozoic extensive faults (De Bellegarde and Gaudenzi, 2002; Nicolai and Gambini, 2007). Finally, during the Plio-Pleistocene period, the chain underwent extensional-transtensional tectonics associated with the opening of the back-arc Tyrrhenian Basin (Casero et al., 1988). The ongoing crustal extension regime contributes significantly to the high seismic activity of this region. This is evidenced by the 1980 Mw 6.9 Irpinia earthquake and the current background seismicity in the area (e.g., Pierdominici et al., 2011; De Matteis et al., 2012).

The interplay between seismicity and circulation of fluids also characterizes the Irpinia region. Geological studies and reports from various sources, such as governmental agencies, academic institutions, and oil and gas exploration companies, have documented the natural fluid occurrences in the area (e.g., Chiodini et al., 2004; Chiodini et al., 2010; Martinelli et al., 2012; Amoresano et al., 2014; Roberts et al., 2017; Gori et al., 2023). The fluid flows can be divided into two systems: the shallower circulation system, which is related to the infiltration of meteoric water, and the deeper circulation system, which depends on the rise of $CO_2$ originating from the mantle. In the shallower system, water infiltrates into the large carbonate karst aquifers, such as the Picentini and Marzano Mountains, and emerges in the Caposele, Capasso and Cassano Irpino main springs at the base of the carbonates (Celico et al., 1979; Allocca et al., 2014; Leone et al., 2021). The deeper system is due to the decarbonatization of the carbonates and is characterized by the rise of $CO_2$, which is temporarily stored in the dolomitic-calcareous porous layers of the middle and upper crust. The $CO_2$ pools entrapped in the Apulian reservoirs can be released through active fault systems. Moreover, the seasonal cycle of the aquifer charging-discharging mechanism seems to control both the micro-seismicity evolution through time at depth and the variation in the amount of gas emissions at the surface (D'Agostino et al., 2018; Chiodini et al., 2020; Buttitta et al., 2023). This is because the shallow hydrological loading forces the opening of deep fractures, leading to an increment of the $CO_2$ degassing process. Additionally, the study area is characterized by hydrocarbon seepages and oil/gas traces at depth, which interact with the deep gas, enriching the complex fluid circulation setting.

## 3. Materials and methods

This section provides an overview of the borehole data used to study the subsurface fluid circulation in the Irpinia area. The focus is on the pore pressures available at the wells. It also describes the procedure for categorizing the geological formations starting from the litho-stratigraphy detected in the well logs. Additionally, the method used for calculating pore pressures from mud weights is explained. Detailed information regarding the data and methodological approaches are found in Supplementary Materials.

### *3.1 Well dataset*

The analysed dataset comprises 13 well logs (Tab. 1 in Supplementary Materials), edited at a scale of 1:1000 and available on the ViDEPI website (ViDEPI database). ViDEPI is an online platform established by the Italian Geological Society, the Ministry of Economic Development, and the Association of Oil Companies active in Italy (Assomineraria) to provide easy access to documents related to expired Italian mining permits and concessions. It currently offers over 2300 well logs from drilling activities in Italy spanning from 1895 to 2016. The selected logs, indicated as "well profiles," are located in the Campania and Basilicata regions and cover the Irpinia area and its surrounding neighbourhoods (Fig. 1a).

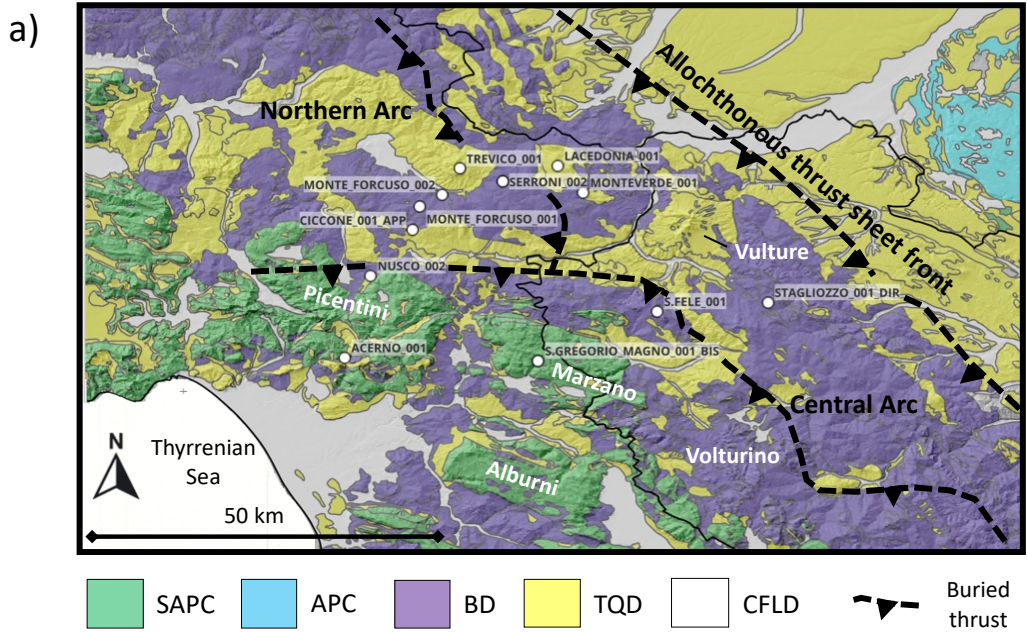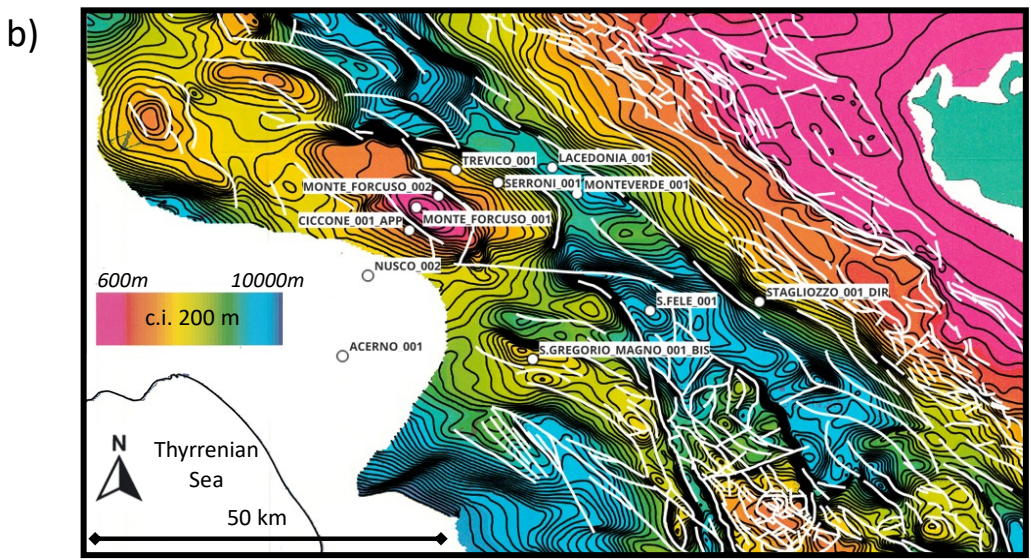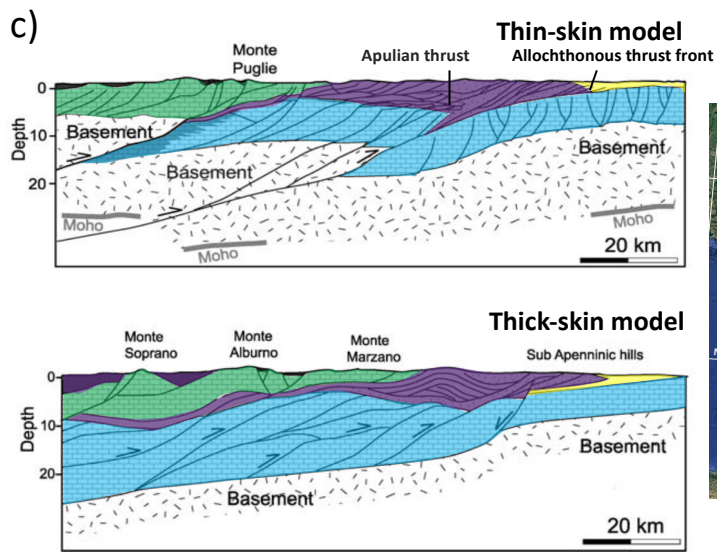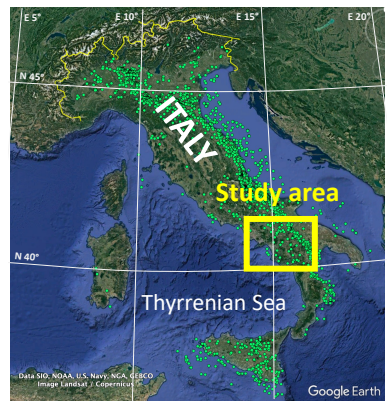

*Figure 1 – Geological maps and sections of the Southern Apennines: a) simplified geological map of Irpinia area (digital elevation model and lithological units are modified from Tarquini et al., 2007 and Italian National Geoportal, respectively); b) depth map of top Apulia Platform and structural lineaments (white lines), modified from Nicolai and Gambini (2007); c) deep structural settings of the Southern Apennines, modified from Scrocca et al. (2005). The maps include the location of the analysed wells (white circles). Legend a) and c): SAPC, South-Apennines Platform Carbonates (light green); APC, Apulia Platform Carbonates (light blue); BD, deep and shallow Basin Deposits (violet); TQD, Tertiary and Quaternary Deposits (light yellow); CFLD, Coastal, Fluvial and Lacustrine Deposits (white) predominantly of Quaternary age. It is worth noting that the depth contours of b) refers to sea level, and the green dots of the Italy base map represent the locations of all the boreholes available on Videpi website (see Section 3.1 for details).*

The analysed wells, drilled between 1957 and 1999, provide a wide range of qualitative and quantitative information, including stratigraphic and lithological descriptions obtained through cutting analysis, geophysical logs (such as resistivity and natural potential logs), gas/oil/water-bearing intervals, percentage of gas detected, geochemical data (chemical and isotopic analyses on fluids), drilling and technical information (such as mud weight, absorption, and loss), core descriptions (lithologies, fractures, petrophysical parameters), production tests (fluid discharge, type, pressure), and the final status of the well (dry, gas, etc.). In this study, all the features related to drilling muds, fluids, and geological series encountered during drilling were collected and integrated to evaluate in-situ pore pressures and fluid flow.

In particular, the pore pressures, which are the pressures of gas or fluids within the porous rock, have been obtained at the well site using both indirect and direct in-situ pressure measurements (Tabs. 1 and 2 in Supplementary Materials). The indirect measurements were estimated from the drilling mud weights, which provide the primary source of information for the selected wells. The direct measures of pore pressure were acquired during fluid formation or production tests, such as the Drill Stem Test (DST), which is used to evaluate the productive capability of a selected rock interval. Although the direct measures are available for a few numbers of boreholes in the study area, these values are consistent with the indirect pressures. Moreover, in some cases, the well profiles provide information on the fracture pressure (known as the Leak Off Test or LOT), which is usually measured for testing the sealing efficiency of a formation after drilling and is obtained converting the mud weight equivalent to the fracking condition into pressure, using a procedure that is described in Section 3.3. Finally, the well profiles indicate the interval affected by total loss of mud circulation, which corresponds to the loss of mud within the drilled rock interval. When this occurs, the mud does not rise to the surface, and therefore no quantitative information on the formation pressure can be obtained.

### *3.2 Categorization of litho-stratigraphy found at the well sites and permeability features*

The successions encountered in the selected wells were grouped in five main categories according to the litho-stratigraphy and structural features provided in the well profiles and the geological knowledge available in literature (e.g., Patacca 2007; Patacca and Scandone, 2007; Pescatore et al., 2008; Torre et al., 2009; Pescatore and Pinto, 2014; Geological Survey of Italy - Sheets 186, 433 and 450). The adoption of a uniform categorization over the whole area, instead of using the multiplicity of formational names presented in the well logs, allow us to enhance the correlations between tectono-stratigraphy, petrophysical features and geo-pressure regimes, as shown in Section 5. The main deposits are listed below:

1) South-Apennines Platform Carbonates (SAPC): the layer comprises lower Triassic-lower Miocene shallow-water calcareous and dolomitic series, deposed in inner platform, platform edge and slope environments. The succession is allochthonous over the area and limited at the base by regional eastward-verging thrust sheets. It is characterised by fair petrophysical properties (medium-high porosity and permeability) due to fracturing, dolomitization and karstification processes (De Bellegarde and Gaudenzi, 2002; Allocca et al., 2014).

2) Apulia Platform Carbonates (APC): the layer includes Mesozoic-Tertiary calcareous-dolomitic deposits belonging to interior platform environment. The sequence, buried over the area and characterised by anticlines and thrust sheets, has locally fair petrophysical features relative to lithofacies texture and fracture network (MICA, 1989; Amoresano et al., 2014). Moreover, local tectonic and diagenetic conditions can lead to the formation of impermeable levels, as in the case of Costa Molina hydrocarbon field (Agri Valley), wherein the impervious Miocene limestones and cemented breccia seal the hydrocarbon-fed carbonates (Mattavelli et al., 1993).

3) Basin Deposits (BD): the layer is constituted by the Mesozoic and Tertiary calcareous and silico-clastic lithologies deposed into both extensional and compressional settings of the South-Apennines area during the pre- and synorogenic phases. It comprises: Cretaceous-lower Miocene units of Sicilide-Liguride Basins; upper Triassic-lower Cretaceous and upper Cretaceous-middle Miocene units of Lagonegrese-Molisano Basin; lower-upper Miocene synorogenic sediments formed within undeformed and deformed foredeep domains. The succession, involved in the fold-and-thrust belt, represents allochthonous sequences sandwiching between the SAPC and the APC layers. With respect to the petrophysical properties, this deposit assumes a more or less permeable behaviour to the fluid flow depending on the lithological content (sandy- or shaly-rich sediments, dolomites, calcareous-sandy deposits) and the diagenetic processes. Evidence of its impermeable behaviour is provided by the numerous springs found at the base of the Southern Apennines karst carbonate aquifers. Indeed, the tectonic contact between the aquifers and the pre- and synorogenic BD controls the groundwater circulation being the deeper hydrogeological boundary of the groundwater-rock system (Allocca et al., 2014). Moreover, its low permeability is confirmed by the petroleum system of the Agri Valley, wherein most hydrocarbon fields are found in the APC reservoirs, sealed on top and laterally by the Miocene marls and shales (Bertello et al., 2004). Conversely, the porosity of the layer increases playing an active role in trapping or flowing fluids, as testified by the petroleum fields related to the Benevento area, where some hydrocarbon-bearing levels belong to the Lagonegrese-Molisano Basin units (Palombi, 1989; Casero, 2005).

4) Upper Tertiary and Quaternary Deposits (TQD): the layer includes upper Miocene-Pleistocene sediments belonging to different depositional environments, such as Messinian evaporites, lower Pliocene shales unconformably laying above APC, and Plio-Pleistocene intra-Apennines successions, which correspond to the sedimentary infilling of the synorogenic wedge-top basins. Due to the occurrences of shale-rich lithotypes and locally gypsum, these successions are characterised by low permeability (e.g., Ingebritsen et al., 2006). In the Agri Valley, the Lower Pliocene clay sequences constitute a good seal for hydrocarbon entrapment (Bertello et al., 2004).

5) Coastal, Fluvial and Lacustrine Deposits (CFLD): these series are predominantly constituted by Pleistocene-Holocene coastal, alluvial and fluvial sediments, deposed under erosion and transport operated by rivers and sea. Due to the high lithological heteropy and the low cementation degree, the layer can be locally considered permeable (e.g., Allocca et al., 2009). This category was not recognised in the analysed wells and thus is not mentioned in the Sections 4.

*3.3 Pore pressure calculation and interpretation*

In the field of oil exploration and production, the estimate of pore pressures from mud weights has been a well-known methodology worldwide (e.g., Dickinson, 1953; Fertl, 1976; Bourgoyne et al., 1991; Bigelow, 1994; Lee and Deming, 2002). The procedure used here is one of the techniques applied to study and map subsurface pore pressures through the analysis of well data and geophysical logs (Zhang, 2011; Radwan et al., 2022). Indeed, the pore pressure prediction plays a key role in ensuring the efficiency and safety of drilling operations, as it directly affects wellbore stability and the risk of blowout or pressure kick.

Usually, the pore pressure at depth is described with respect to the hydrostatic pressure (e.g., Zoback, 2007). Hydrostatic pressure increases continuously as a function of depth and water column weight (from the surface to the depth of interest). The fundamental relation is:

$$P = \int_0^z \rho w(z) g \, dz \approx \rho w g z \quad (1)$$

where $P$, in the case of hydrostatic pressure, is the water pressure at the burial depth $z$, $\rho w$ is the water density (1 g/cm$^3$, in the case of pure water) and $g$ is the gravitational acceleration (9.81 m/sec$^2$).
This formula can also be used to calculate the lithostatic pressure and the pore pressure mud-weight-derived, by changing the fluid density. Indeed, the lithostatic pressure increases with depth and the weight of the overburdened rock column, which usually has a mean density of 2.3 g/cm$^3$ in sedimentary basins. On the other hand, pore pressure at the well-site is the weight of the fluids in the formation encountered while drilling, balanced by mud. The pore pressure derived by mud weight, called "mud-derived formation pressures" from here on, is the upper limit of formation pressure because the mud weight is commonly set about 10% higher than the rock pressure to prevent the hole from collapsing. Hydrostatic pressure is also known as normal pressure, and pressures that deviate 1MPa from it are called anomalous or abnormal pressures (equivalent to 100 m of hydraulic head; cfr. Neuzil, 1995). Positive pressures that are higher than the hydrostatic are called overpressures, while negative pressures that are lower than hydrostatic are called underpressures. All these pressure types are illustrated in the pressure-depth plot shown in Figure 2. It's important to note that in the regions characterised by a normal fault regime, such as the Irpinia area, the overburden rock density integrated from the surface to the depth of interest also corresponds to the vertical stress, which is the maximum principal stress. The least principal stress is below the vertical one (horizontal stress). LOT measures mentioned in Section 3.1 provide information about the minimum horizontal stress and are crucial in determining whether the measured fluid pressures are close to fracturing the sealing zone, leading to fluid loss (leakage risk in Fig. 2).
Finally, the last relation useful to compare overpressure values at different depths defines the dimensionless normalized pressure ($Pn$) as:

$$Pn = \frac{Pf - Ph}{Pl - Ph} \quad (2)$$

where $Ph$, $Pl$ and $Pf$ correspond to the hydrostatic, lithostatic and pore pressure, respectively (e.g., Lee and Deming, 2001). $Pn$ ranges from hydrostatic pressure (equal to 0) to lithostatic pressure (equal to 1).

The conceptual framework on the pore pressure calculation presented above can be applied whether the pore spaces from the surface to the depth of interest are interconnected or not. Both these hydraulic conditions represent commonly-observed scenarios, even in the Irpinia area and surrounding regions.

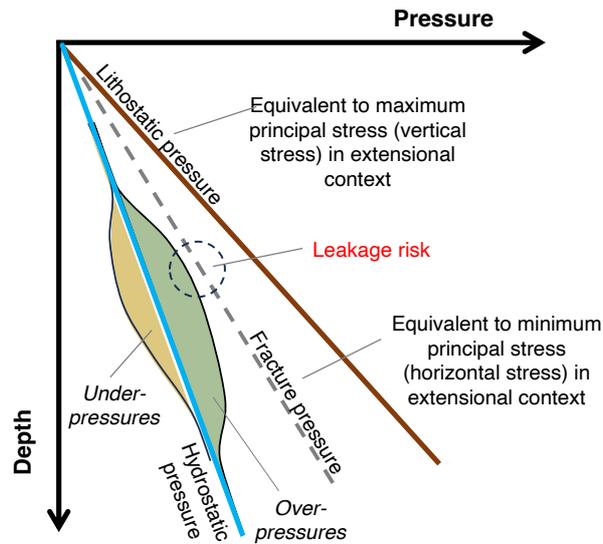

*Figure 2 – Conceptual pressure-depth plots to describe the pore pressure regimes measured or calculated at the well sites.*

Figure 3 shows two hydraulic scenarios: continuous (Fig. 3a) and discontinuous (Figs. 3b-3c) interconnection of the pore fluids along the depth. All proposed sketches show water-filled sandy and shaly sequences that correspond to permeable and impermeable layers, respectively.

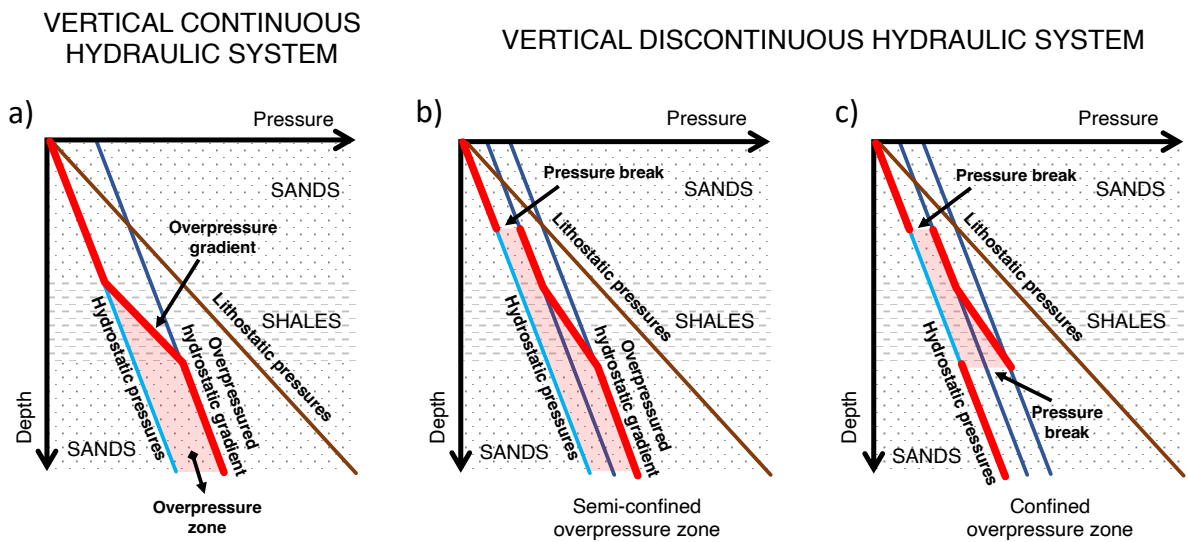

*Figure 3 – Notions and conceptual schemes to describe the vertical pore pressure regimes measured or calculated at the well sites.*

Additionally, the shaly series is assumed to be under the action of tectonic stresses or loading processes and thus generates overpressures. In all sketches, the porous layers are characterized by hydrostatic pressures and overpressures that align over hydrostatic gradients (i.e., overpressured hydrostatic gradients). In the case of a vertical hydraulic continuity, the overpressured hydrostatic gradients are reached for pressure transferring from the pressure generation zone, as the shaly layers, to the permeable zone, as shown in Fig. 3a and downward of Fig. 3b. In the case of a vertical discontinuous hydraulic system, upward of Fig. 3b and Fig. 3c show that the top permeable layer is characterized by a pressure break, separating hydrostatic pressures from overpressures over the hydrostatic gradient. Depending on whether a pressure break occurs at the base of the pressurized interval, the overpressure zone can be considered a semi-confined or a confined interval (Figs. 3b-3c).

The examples shown in Figure 3 do not deal with all the possible hydraulic connections occurring through the tectonostratigraphic contacts (e.g., Watts, 1987; Yielding et al., 1997), but highlight the processes of pressure generation and transmission through fluid flow. As mentioned in Section 1, it is worth noting that the time factor is a crucial element in defining the hydraulic behaviour of a rock system, both because the formation of overpressures can occur through physical processes that are confined in time, as in the case of hydrocarbon generation (e.g., Welte et al., 1997), and because the transfer processes change over time, as in the case of sealed faults that later turn into leaking fault zones (e.g., Gudmundsson, 2001).

Finally, in the wake of the mechanisms controlling the overpressure dissipation in shale-rich successions, as in the case of the Irpinia area, some authors prefer to discuss the hydrodynamic conditions of a rock system using a diagram of density-adjusted hydraulic head versus depth (e.g., Neuzil, 2015). We also adopt this type of diagram to discuss the geological forcings in relation to the observed overpressures. The variation of hydraulic head ($\Delta h$) is computed according to the following relation:

$$\Delta P = \rho w g \Delta h \quad (3),$$

where, $\Delta P$ is the difference between pore pressures and is calculated using (1) at different depths ($z$), $\rho w$ is the water density (1 g/cm$^3$, in the case of pure water) and $g$ is the gravitational acceleration (9.81 m/sec$^2$).

**4. Data analysis**

In this section, the mud-derived formation pressures, calculated and normalised according to the methodology described in Section 3.3, are presented well by well and in relation to the geological formations and gas occurrences. Computed values are displayed by means of pressure-depth plot, in which hydrostatic and lithostatic pressures, calculated according to 1.0 g/cm$^3$ for the water density and 2.5 g/cm$^3$ for the bulk density, are also superimposed. The former value generally refers to pure water, while the latter corresponds to an average value of bulk densities obtained for the stratigraphic units of the Southern Apennines (cfr. Montone and Mariucci, 2023). The stratigraphic schemes of the selected borehole are also added to these profiles for better evaluating the correlation between pressures and geologic boundaries. All these graphical products are grouped according to the total depth reached by the boreholes and presented in Figure 4, while complete information on mud density, mud circulation loss, leak-off test, production test pressure and stratigraphy may be found in Tables 1 and 2 of Supplementary Materials. Finally, the data analysis focuses on the relationship between the overpressures or the hydraulic head variations and the fluids found in the wells analyzed. Fluid occurrences and their chemical features have been collected from the well profiles and are listed in Table 3 of Supplementary Materials.

*4.1 Mud-derived formation pressures at well sites*

Most of the selected wells show continuous pressure regimes close to the hydrostatic gradient (Figs. 4a-4f) or in slight overpressure gradient (Figs. 4i and 4l). The exceptions are the Acerno 1 and Ciccone 1 wells, which exhibit the highest overpressure gradients (Figs. 4g and 4h).

In particular, the pressure profile of Acerno-1 is characterised by a hydrostatic gradient within the South-Apennines platform deposits and in the upper part of the Lagonegrese basin unit, up to 1993 m of depth (boundary b1 in Fig. 4g). Downward, it is followed by a confined interval of strong overpressures, up to 4439 m of depth (boundary b2 of Fig. 4g). The upper boundary b1 corresponds to a tectono-stratigraphic contact within the Lagonegrese basin unit, while the lower boundary b2 occurs within the shallow-water carbonates of the Apulia platform (Patacca, 2007), without apparent variations of litho-stratigraphic or tectonic features. Below the pressure break b2, pore pressures turn to the hydrostatic gradient.

The pressure profile of Ciccone-1app shows a continuous increase in positive pressure within the Oligo-Miocene allochthonous unit, up to a depth of 2426 m (boundary b1 in Fig. 4h). The main growth of overpressures occurs within the shale-rich formations below 1404 m depth. Moreover, the semiconfined overpressure interval is limited at the bottom by a b1 pressure break located very close to the level of the Pliocene limestone breccia. Below the b1 limit, the pressure regime is unreliable due to the recurrence of total losses in the mud circulation.

Among the shallow boreholes, Monte Forcuso-1 and Monte Focruso-2 exhibit continuous pressure profiles close to the hydrostatic gradient (Figs. 4a and 4b). In addition, the low-pressure values calculated using the mud weights at Monte Forcuso-1 are also confirmed by direct pressure measurements acquired during two formation tests.

Although S. Gregorio Magno-1bis and S. Fele-1 cross different tectono-lithostratigraphic units, these two wells show similar pressure regimes (Figs. 4i and 4l). In fact, S. Gregorio Magno-1bis is characterised by hydrostatic gradients within the deposits of the South-Apennine platform and in the upper part of the Lagonegrese basin unit (see also Patacca, 2007), up to a depth of 2600-2700 m. Downwards, the pressure increases, reaching slight overpressures at the bottom of the borehole. S. Fele-1, which crosses lower Triassic-lower Miocene basin deposits (see also Patacca, 2007), shows a hydrostatic gradient up to a depth of 2300 m and slight overpressure gradients at deeper levels. About S. Fele-1, it is worth noting that the mud weights used to calculate its overpressures are of poor quality because refer to a range of density values and, especially, below 2300m of depth, the ranges of densities are very broad. The overpressures shown in the Figure 4l are computed using the maximum density from the available density ranges at each depth and are likely overestimated from 2300m to the bottom of the borehole. Additionally, the densities refer to considerable layer thicknesses, and hence, the overpressures of thin layers are likely to be underestimated and not visible.

Figure 4m shows the maximum normalized pressures (Pn) at the well sites and includes the available fracking pressures of the basin deposits (BD) collected from the Leak Off Test (LOT) of the Monte Foi-1 (ViDEPI database). Although this borehole is not analysed in the present study, it shows similarities with S. Fele-1 regarding structural setting and drilled basin deposits. The plot reveals that the highest pore pressure values are measured at Acerno-1 and Ciccone-1 and that even the maximum Pn is reached by the basin deposits of S. Fele-1, these values are still far from the fracking conditions.

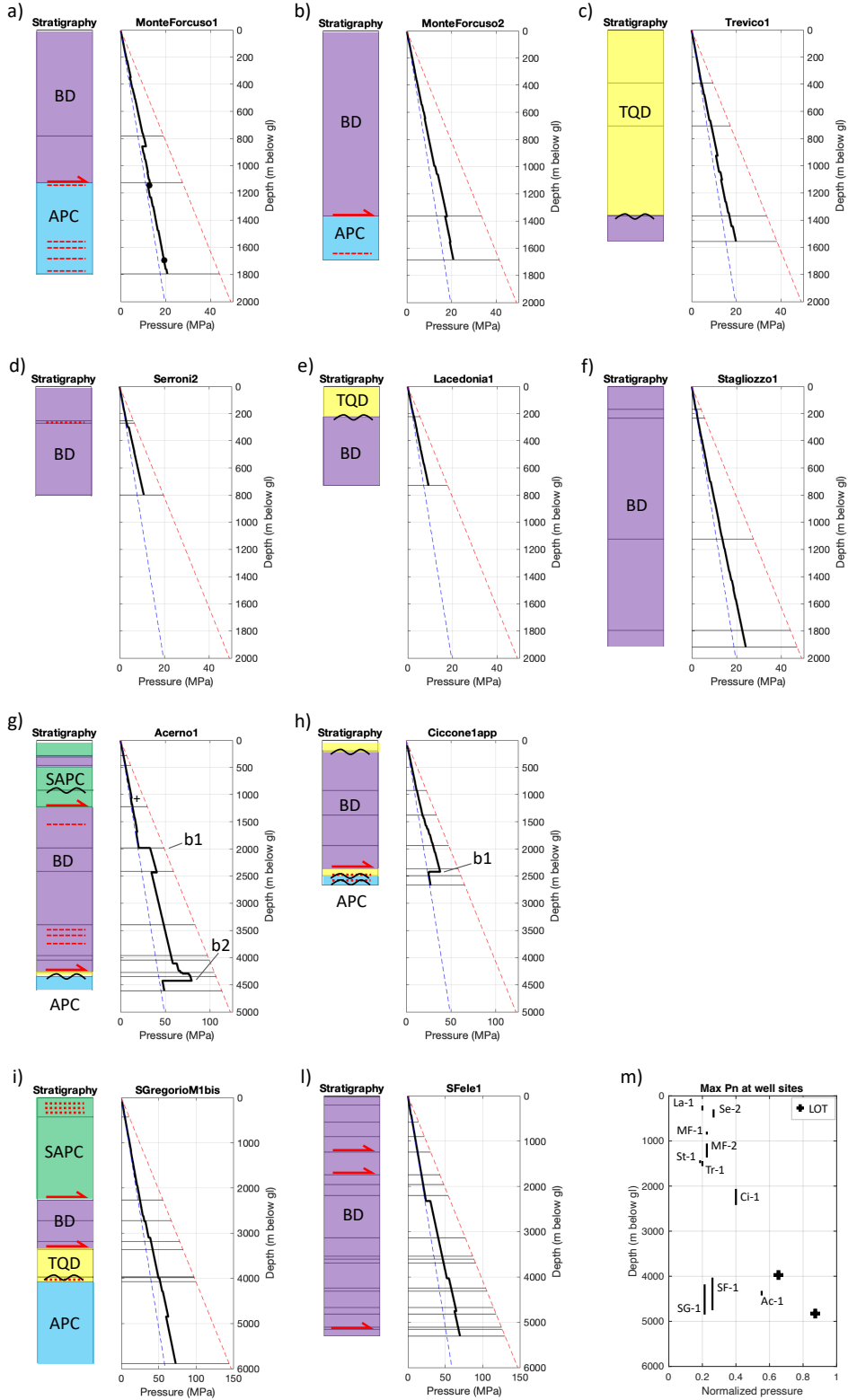

*Figure 4 – Mud-derived formation pressure profiles of the selected wells. The pressure-depth diagrams are paired with the stratigraphic schemes of the wells and refer to a) Monte Forcuso-1, b) Monte Forcuso-2, c) Trevico-1, d) Serroni-2, e) Lacedonia-1, f) Stagliozzo-1dir, g) Acerno-1, h) Ciccone-1app, i) S. Gregorio Magno-1bis and l) S. Fele-1. Note that in the pressure-depth plots the pressure (x-axis) varies as a function of the depth (y-axis), which reaches 2000 m in graphs a)-f), 5000 m in graphs g)-h) and 6000 m in graphs i)-l). Plot m) shows the maximum normalised pressure (Pn) at the well sites and includes the fracking pressures of the basin deposits (BD) collected from the available LOT (see also the text for details). The well names are indicated with the abbreviations: MF-1 and MF-2 (Monte Forcuso-1 and -2), Tr-1 (Trevico-1), Se-1 (Serroni-2), La-1 (Lacedonia-1), St-1 (Stagliozzo-1dir), Ac-1 (Acerno-1), Ci-1 (Ciccone-1app), SG-1 (S. Gregorio Magno-1bis) and SF-1 (S. Fele-1).*

Finally, the calculated normalised pressures are grouped into four hydraulic regimes ranging from hydrostatic or near-hydrostatic conditions to overpressure gradients of increasing magnitude (slight, moderate and strong overpressures). Figure 4m depicts the maximum normalised pressures and highlight, as mentioned above, that most of the selected wells reach overpressure gradients below 0.3 (slight overpressure gradients), with the exception of Ciccone-1app and Acerno-1, which reach 0.4 (moderate overpressure gradients) and 0.55 (strong overpressure gradients), respectively. Table 1 reveals with more precision the depth intervals in which the pressures of the analysed wells fall into the proposed pore pressure clusters.

| Well name | HyC (< 0.15) NHyC (0.15-0.2) | SLOG (0.2-0.3) | MOG (0.3-0.5) | STOG (> 0.5) |
|---|---|---|---|---|
| Acerno-1 | 0-2000 m; 4427-4613 m | - | 2000-4296 m | 4296-4427 m |
| Ciccone-1app | 0-922 m | 922-1574 m | 1574-2420 m | - |
| Lacedonia-1 | 0-730 m | - | - | - |
| M. Forcuso-1 | 0-225 m; 857-1796 m *225-760 m* | 760-857 m | - | - |
| M.Forcuso-2 | 0-86 m *86-1056 m; 1366-1686 m* | 1056-1366 m | - | - |
| S. Fele-1 | 0-2321 m | 2321-5306 m | - | - |
| S. Gregorio M.-1bis | 0-3100 m *3100-4180m; 4850-5890m* | 4180-4850 m | - | - |
| Serroni-2 | 0-299 m | 299-799 m | - | - |
| Stagliozzo-1 | 0-1192 m *1192-1919 m* | - | - | - |
| Trevico-1 | 0-666 m; 926-1346 m *666-926 m; 1346-1557m* | - | - | - |

*Table 1 – Depth intervals and magnitude of normalised pressures (Pn) calculated at the selected well sites. The magnitude of the normalised pressures is classified using four intervals: HyC and NHyC, hydrostatic and near-hydrostatic conditions, respectively; SLOG, slight overpressured gradients; MOG, moderate overpressured gradients; STOG, strong overpressured gradients. It's worth noting that the depth is below ground level and the line in italics refers to NHyC.*

*4.2 Overpressures and fluid occurrences at well sites*

All wells, except Lacedonia-1 and Trevico-1, have small amounts of hydrocarbon-related gas occurrences. Some wells have traces of fresh, brackish and saltwater, while others have traces of oil and levels or seepage of $CO_2$.

In particular, Acerno 1 Well shows low amounts of natural gas produced by thermogenesis (< 4%), found above and within the over-pressurised interval, which is confined between b1 and b2 hydraulic barriers (Fig. 5a). These gas pockets are mainly concentrated in the low permeability strata, i.e., the shaly and silty sequences of Lagonegrese Basin units (BD). In addition, during the formation test (Drill Stem Test 1, or DST-1) performed at the bottom of the Apulian carbonate formation (APC), carbon dioxide with $H_2S$-associated (>90%), mud, connate water and oil (even in traces) were extracted. The isotopic analysis of carbon dioxide mentioned in the well profile indicates that such gas is originated from the thermal degradation of the carbonates, as well as the main $CO_2$ earth degassing of the southern Apennines (Chiodini et al., 2004; Buttitta et al., 2023). Moreover, although the detection analysis used for discriminating methane and superior hydrocarbon compounds was contaminated by oil, light hydrocarbons originated by thermogenesis are also mentioned in the well log at this level. No free groundwater levels are indicated in the well profile, which likely confirms the impervious texture of the entire interval. As mentioned, the only occurrence of paleo-water is below 4433m of depth (DTS-1).

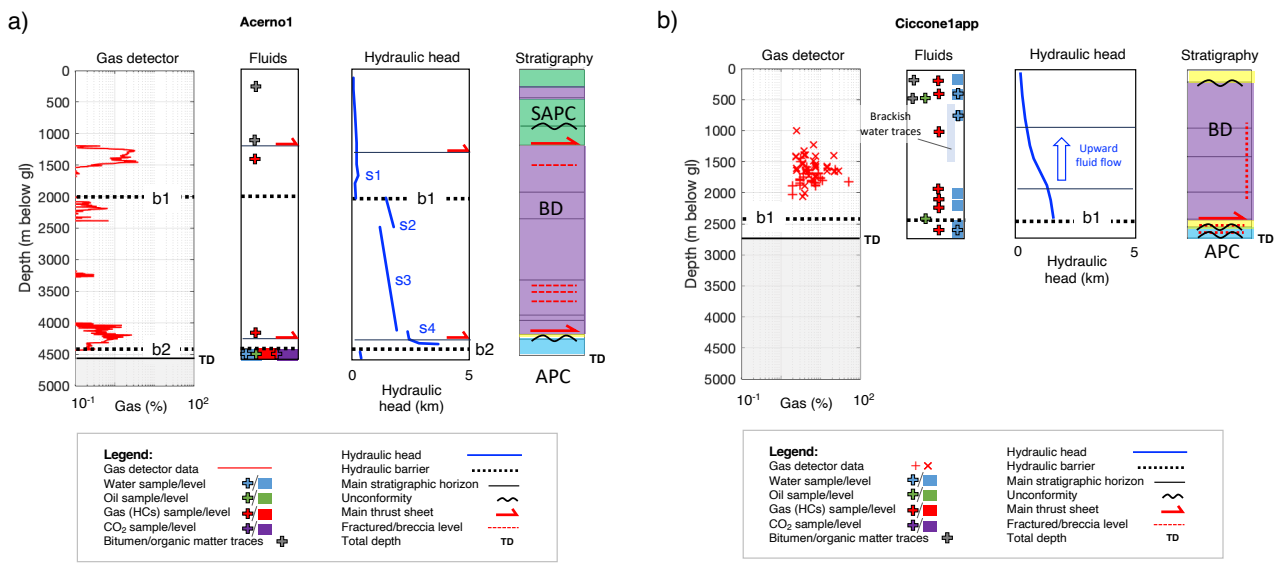

*Figure 5 – Hydrocarbons, $CO_2$ and groundwater occurrences in: a) Acerno-1 and b) Ciccone-1app. Plots depicting gas detector data and fluids are displayed with hydraulic head and stratigraphic information.*

The occurrence of gas peaks at certain depth intervals is not directly connected to a relevant variation of hydraulic heads (segment s1-s4 in Fig. 5a), by means of overpressures. This is because the depth intervals of gas peaks correspond both to hydrostatic conditions (hydraulic head segment s1) and overpressure gradients (segments s2, s3, and s4).

In Ciccone-1app, natural gas traces were found without $CO_2$ (cfr., De Bellegarde and Gaudenzi, 2002). The gas detector tool recorded scattered amounts of gas in the intensively fractured layers between 930 to 2110m of depth (Fig. 5b). The highest gas values were found in the light-dark grey shales with thin intercalations of sandstone and siltstones between 1383 and 1944m of depth. This layer is identified as the Tertiary basin units (BD) and is considered a low permeability interval, as confirmed by the lack of free groundwater levels in the borehole. At these depths, there is a rise in overpressure, as illustrated in Figure 4h, and confirmed by the significant growth of the hydraulic head. This evidence indicates a relationship between overpressure gradients and the occurrence of gas in the shaly sequences. The overpressures are constrained below by the hydraulic barrier b1, corresponding to the Pliocene calcareous breccias above the APC, and disappear within the bottom hole sequence.

In Serroni-2, the shaly interval with gas traces recorded below 300m of depth corresponds to the slight overpressure gradient zone (Fig. 6a). Moreover, the gas trace contains methane (~95%), hydrocarbon-superior molecules (~4%), and nitrogen (~1%). Although Serroni-2 has a lower total depth and a lower overpressure gradient compared to Acerno-1 and Ciccone-1app, it reaffirms the correlation between overpressures and low permeable deposits filled with low concentrations of light hydrocarbons (predominantly methane).

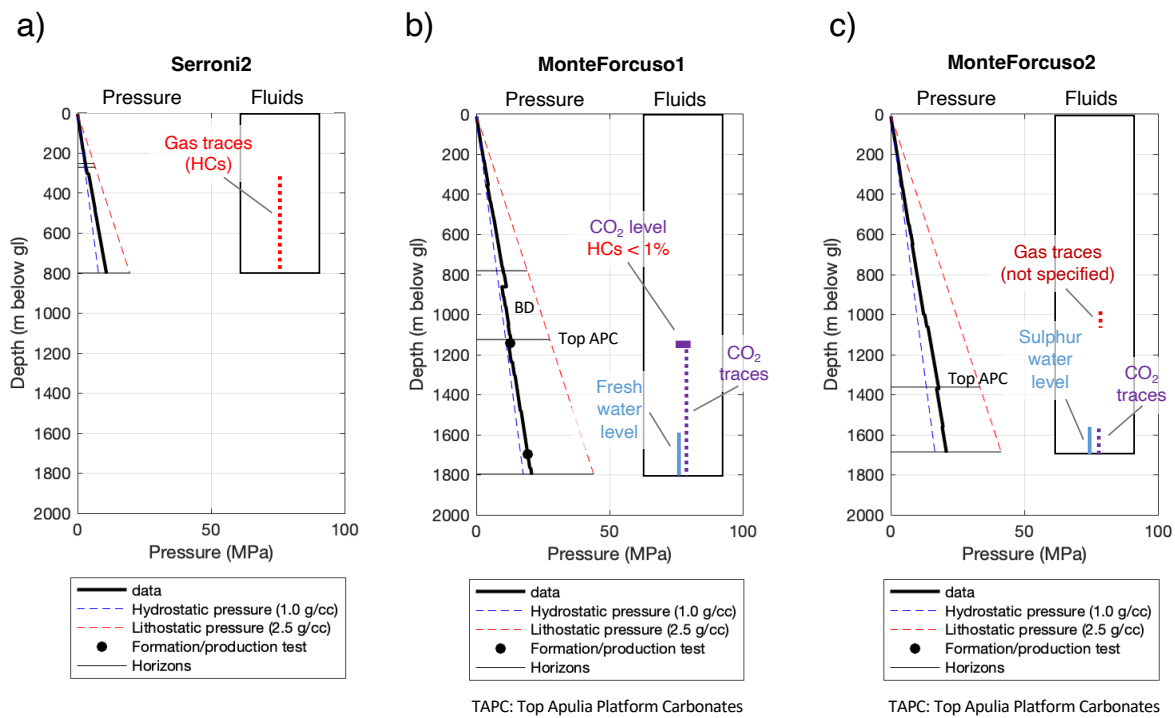

*Figure 6 – Pressure trends and fluid occurrences (hydrocarbons, $CO_2$ and groundwater) in some of the selected wells: a) Serroni-2, b) Monte Forcuso-1 and c) Monte Forcuso-2. Note that BD is for Basin Deposits and APC is for Apulian Platform Carbonates.*

Gas occurrences have been discovered in the shallow-water carbonates (APC) of Monte Forcuso-1, according to the well profile (Fig. 6b). The gas is present throughout the sequence from depths of 1128m to 1800m (bottom hole). The mineralization column of the well log displays a $CO_2$-bearing level between depths of 1165m and 1128m. This level was analysed in detail and two gas samples showed that the gas is primarily composed of carbon dioxide (96.7-99.7%), nitrogen (0-2.4%), methane (0.3-0.9%), with some traces of superior compounds of light hydrocarbons and $H_2S$. The gas fills the intensely fractured Senonian carbonates and is sealed by the silty marls and shaly sandstones above (BD). The Core 2, extracted at about 1052m of depth within silty marls and sandy deposits, exhibits a poor permeability and a porosity not fed by gas. Furthermore, the mineralization column shows traces of $CO_2$ below 1165m and up to the bottom hole, where a freshwater level 200m thick also fed the Neocomian carbonates and dolomites. The slight overpressure gradient in the BD is unrelated to the gas occurrence because the mineralization is limited to the APC. Finally, the Mesozoic layer is also found in hydrostatic conditions, confirming that overpressure and gas occurrence are uncorrelated at the well site.

Monte Forcuso-2 exhibits traces of carbon dioxide in the sulphur weakly salty water level at the base of the boreholes, between 1551m and 1691m of depth. The geological notes of the well profile indicate the presence of a low permeability stratum covering the carbonate reservoir, made up of shaly and marly arenaceous deposits. The annotations also suggest that some more permeable arenaceous levels in the upper part of the silico-clastic sequence contain very low amounts of gas (e.g., around 1000m of depth). Although the gas composition is not specified, this gas occurrence matches with a slight overpressure increment observed at the same depth in the pressure profile, confirming the relation among shaly lithotype, gas traces, and overpressures.

S. Fele-1, which is one of the deepest boreholes being analysed, doesn't show any significant correlation with the fluid traces discovered during drilling as illustrated in Figure 7a. This result is likely unreliable because the overpressures are likely overestimated below 2300m of depth to the bottom of the borehole, as already described in Section 4.1. Additionally, the densities used for calculating pore pressures refer to considerable layer thicknesses, and hence, the overpressures of thin layers or in proximity to the thrust sheets are likely to be underestimated and not visible.

Finally, in S. Gregorio Magno-1bis, the deepest borehole of the study area, an increase in overpressures is noted in the more pelitic sequences containing a small amount of organic matter (Fig. 7b).

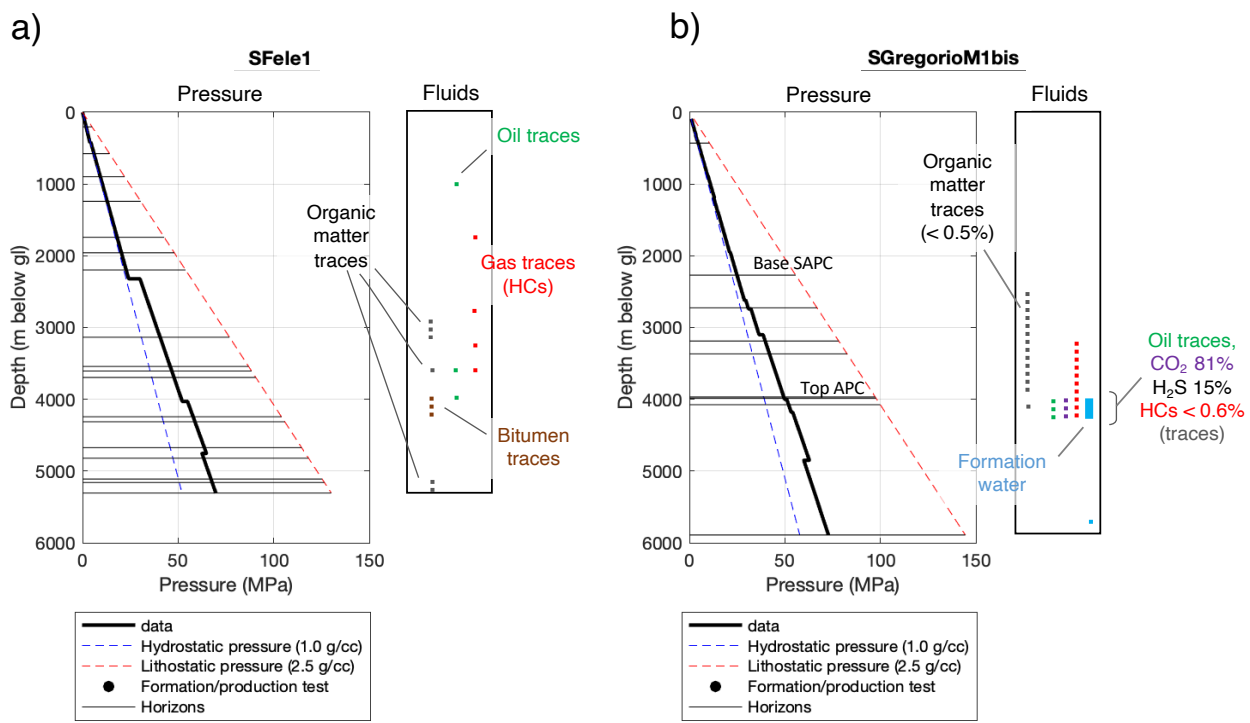

TAPC: Top Apulia Carbonate Platform
BSAPC: Base South-Apennines Platform Carbonates

*Figure 7 – Pressure trends and fluid occurrences (hydrocarbons, $CO_2$ and water) in some of the selected wells: a) S. Fele-1 and b) S. Gregorio Magno-1bis. Note that APC is for Apulian Platform Carbonates and SAPC is for South-Apennine Platform Carbonates.*

## 5. Discussion

This Section aims to discuss the processes that led to the formation of overpressures in low-permeability sequences found in the analysed wells. As per the previous sections, the overpressures are not related to the age of the tectonostratigraphic units crossed in the well site. This is because slight to strong overpressured gradients are found in both the Mesozoic and Cenozoic deposits. Also, the facies deposited in specific paleogeographic domains do not correlate with the overpressure values as higher overpressures are observed in both the pelagic basin deposits and the foredeep deposits involved in the chain structuration. Apart from the causes related to low-permeability lithologies, which may hinder fluid movement and locally generate overpressures, and associated with the presence of trace gases like methane in clay lithologies, which may lead to small gradients of overpressures, we will examine whether geological forces such as tectonic stresses have been active over a long period, for example, hundreds of thousands or millions of years, in low-permeability strata. To achieve this, we present additional petrophysical and hydraulic parameters and discuss the origin of the highest overpressures found in Ciccone-1app and Acerno-1 (ranging from moderate to strong overpressured gradients).

In particular, in the Irpinia area, the structuration of the Apennines chain as a fold-and-thrust belt was made up of a compressional phase from the upper Tortonian to the middle Pleistocene (10-0.5 million years), followed by an extensional phase that is still ongoing (e.g., Improta et al., 2003). These tectonic mechanisms may have affected the shaly rocks, leaving a footprint of natural and slow transient pressure-disturbing mechanisms (Neuzil, 2015). The rock system under the action of tectonic forces can be described using the transient Darcy flow model, which suggests that in non-stationary (transient) conditions, the inflow and outflow of a specific rock volume may not be zero due to the compressibility of the rock and fluid (e.g., Toth, 2009). The external tectonic force can also affect the fluid-rock system by partially changing its porosity and pore pressure (Neuzil, 1995). Moreover, the stress acting, and the relative volumetric strain can be expressed through time or in terms of rate. To investigate the nature of perturbations in the study area and quantify the influence of tectonic stress on the well sites, we consider the geological forcing and the timing of dissipation. The geological forcing ($\Gamma$) refers to the mechanism capable of producing abnormal pressures (expressed in temporal rate), while the timing of dissipation (t) refers to the time it takes for half of the forcing magnitude to dissipate. These aspects can be visualized along with the geometric and hydraulic parameters of the flow domain (i.e., hydraulic conductivity or K, specific storage or Ss, half-length of flow domain or L), as shown in Figure 8.

According to the data presented in Table 2 and Figure 5b, Ciccone-1app shows a significant increase in hydraulic head in the Oligocene and Miocene flysch formations. These formations primarily consist of shales (BD) and are located at a depth of 900 to 1900 meters. The hydraulic head increases from 0.3 km to 1.1 km, indicating a movement of fluids upwards.

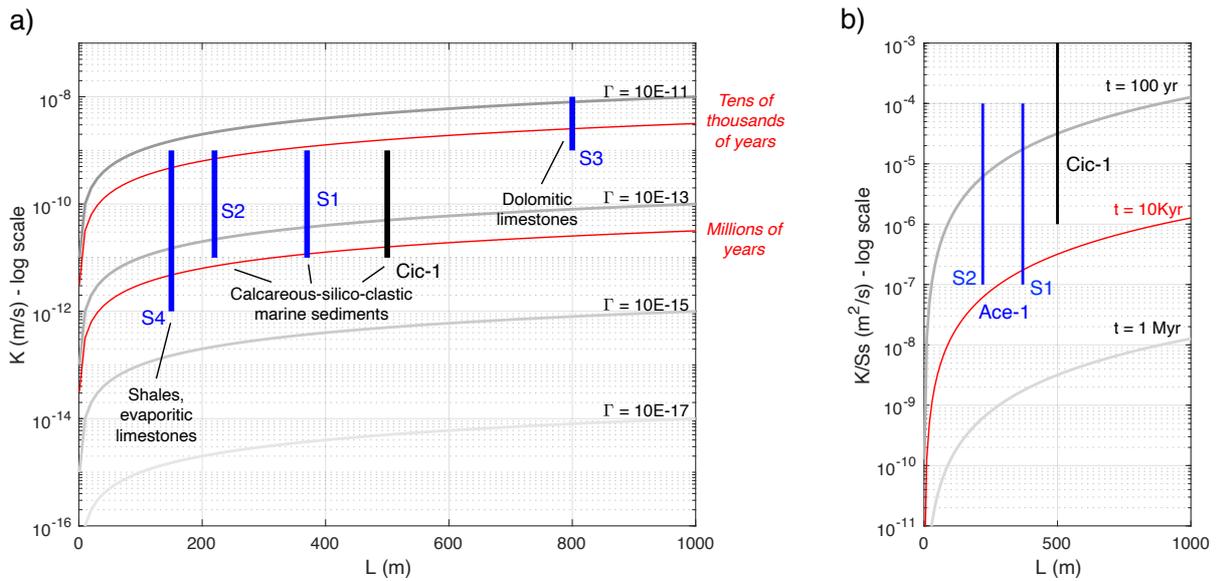

Figure 8 – Comparison between geological forcing and hydraulic parameters of the flow domain at Acerno 1 and Ciccone 1app well sites. The figure shows the timing of perturbation generation (Γ) and dissipation (t). It's worth noting that in figure a), tags S1-S4 refer to Acerno 1 well (see also Fig. 5a and Tab. 2). Also, in figure b), Myr and Kyr are abbreviations of millions and thousands of years, respectively.

In addition, the multiple groundwater levels and water traces in the borehole indicate the presence of free water occurrences that could be moved. Moreover, although sidewall and bottom hole cores are available for the crossed formations, their petrophysical properties have not been incorporated in the well profile. Therefore, analogue case studies are used, and, in particular, the hydraulic conductivity of the flysch sequences is retrieved from outcropping formations studied in the Irpinia, Sannio, and Daunia areas (Bonanno, 2020; Losacco et al., 2021; d'Onofrio et al., 2023; see also Table 3 in Supplementary Materials).

| Well name | Layer | Depth interval (m) | Head variation (km) | Hydraulic conductivity K (m/s) | Specific storage Ss (1/m) |
|---|---|---|---|---|---|
| Ciccone-1 | BD | 900-1900 | 0.8 | $10^{-11} - 10^{-9}$ | $10^{-7} - 10^{-5}$ |
| Acerno-1 | BD | 1230-1990 (s1) | 0 | $10^{-11} - 10^{-9}$ | $10^{-6} - 10^{-4}$ |
| Acerno-1 | BD | 1990-2440 (s2) | 0.3 | $10^{-11} - 10^{-9}$ | $10^{-6} - 10^{-4}$ |
| | BD | 2440-4100 (s3) | 0.7 | $10^{-9} - 10^{-8a}$ | |
| | BD | 4100-4400 (s4) | 1.7 | $10^{-12} - 10^{-9}$ | |

Table 2 – Hydraulic properties of the analysed formations. Note that the hydraulic conductivity denoted by superscript 'a' is derived from a permeability range measured on an analogue rock sample (for the conversion factors see also Freez and Cherry, 1979).

Specific storage is attributed based on published data for shaly and clayey sediments outcropping in different parts of the world (Neuzil, 2015). According to this information, the fluid domain of the shale-rich layers can be characterized by hydraulic conductivities ranging from $10^{-11}$ and $10^{-9}$ ms$^{-1}$ and a flow half-length of 500m (Fig. 8a; Tab. 2). According to these values, the geological forcing that can develop overpressures varies between $10^{-14}$ and $10^{-12}$ s$^{-1}$, i.e., from tens of millions to hundreds of thousands of years and is comparable to the mentioned tectonic stress acting on the Irpinia area in the last ten million years. The specific storage is attributed to the flysch formations with uncertainty since it refers to sites that differ in age and depositional environments from the shaly layers of the study area. It ranges between $10^{-7}$ and $10^{-5}$ m$^{-1}$ (Tab. 2). According to this indication, the dissipation time results to be less than tens of thousands of years (Fig. 8b), matching with a geological strength that has been exhausted over thousands of years, by means during the Holocene period.

Regarding Acerno-1, three hydraulic head segments are recognized in the basinal sequences (BD) bounded by the hydraulic barriers b1 and b2 (Fig. 5a; Tab. 2). The upper segment, indicated with S2 in Figure 5a, is 450m thick and corresponds to shaly marls and shales (Flysch Galestrino and Scisti Silicei formations). The middle segment S3 is thicker than 1600m and mainly refers to dolomitic limestones (Calcari con Selce and Monte Facito formations). The lower segment S4 is 300m thick and corresponds to shales and evaporitic limestones (Flysch Galestrino and Messinian Evaporites). Above b1, more than 750m of shales, marls and limestones (Flysch Galestrino, Scisti Silicei, and Calcari con Selce formations) are also found in hydrostatic conditions (S1 in Fig. 5a). The hydraulic conductivity of these lithotypes has been estimated by considering the available petrophysical data from a sample of dolomitic limestones recovered at 3800m of depth in S. Fele-1. The sampled permeabilities have been suitably converted into hydraulic conductivity (Freez and Cherry, 1979) and used as representatives of the dolomitic limestone interval (S2), as well as attributed to the other levels (S1, S3, and S4) modifying its amount based on the shale content of the analysed layers. The smallest values are attributed to the evaporite lithotype, as also confirmed by literature (e.g., Ingebritsen et al., 2006). Concerning the specific storage, it is collected from analogues of Mesozoic marine formations composed of argillite and marls and outcropping in France (e.g., Boisson et al., 2001). Based on lithotype similarities, it is attributed even with uncertainty only to the shaly and marly sequences (S1 and S2).

Except for S3, all the values plotted in Figure 8a indicate that the geological strength capable of generating overpressures varies from tens of millions to hundreds of thousands of years and is comparable to the tectonic stress acting on the Irpinia area. Hydraulic head segment s3 is influenced by a more recent geological strength, occurring from tens of thousands to thousands of years. Furthermore, among the segments S1-S4, only S1 shows no variation in the hydraulic head, meaning that there is no fluid flow (Tab. 2). This can be explained by considering that the shallow formations are in hydrostatic conditions up to 2000m of depth (SAPC and BD in Figs. 4g and 5a). The fracture gradient is very close to the hydrostatic trend, as indicated by the LOT measurement available for the shallow water carbonates at 1070m of depth (see also Table 2 in Supplementary Materials). Thus, a few increments in the hydraulic head can generate fracking and fluid flow, dissipating the overpressures. Additionally, it is worth noting that the well profile indicates total loss of mud circulation at 1700m of depth, which coincides with a limestone fractured level, thus confirming the occurrence of a depletion point in the pore pressures and the weak resistance of the rock to fluid pressure. Finally, according to the attributed specific storage values, the dissipation of the force magnitude is comparable to a period of tens of thousands and thousands of years (Fig. 8b), by means during the upper Pleistocene-Holocene period.

**Conclusions**

A study was conducted in the Irpinia seismic area to investigate subsurface pore pressures using exploration well data. The study was able to reconstruct the vertical pore pressure trends up to nearly 6km of depth and could potentially be used to calibrate fluid pressure regimes inferred by other geophysical studies at the regional scale. The study also examined the relationship between pore pressures, subsurface geological formations, and fluid occurrences such as hydrocarbons and carbon dioxide. It also addressed the possible origin of the observed overpressures.

The analyzed well profiles show that the majority of the wells exhibit hydrostatic, near-hydrostatic conditions, and slight overpressured gradients in the Meso-Cenozoic platform-basin deposits and Tertiary foredeep flysch sequences. Locally, moderate overpressured gradients are shown in the Triassic-lower Cretaceous pelagic layers (Acerno-1) and the Tertiary flysch deposits (Ciccone-1app). The highest overpressures are found in Messinian evaporites and lower Pliocene shales (Acerno-1 and Ciccone-1app). Thus, specific age of deposits and paleo-depositional environments do not exclusively control the positive pressures.

Regarding the relation between overpressures and fluid occurrences, the study found that the slight overpressures in shallow layers are primarily observed in shaly lithologies that contain traces of gas, mainly hydrocarbons. At the same time, the moderate and strong overpressured gradients at great depths are not correlated to gas occurrences. Still, they are usually associated with or limited by fractured and overthrust zones, such as thrust sheets.

Finally, the study examined the timing of the geological forcing that generated the overpressures. The geological strength capable of generating overpressures varies from tens of millions to hundreds of thousands of years and is comparable to the compressional tectonic stress that was active in the past on the Irpinia area. Moreover, the dissipation of the geological forcing is similar to a period of tens of thousands and thousands of years, like the upper Pleistocene-Holocene period.

**Acknowledge**

We thank Dr. Mauro Buttinelli of the Department of Seismology and Tectonophysics (INGV) and Prof. Simonetta Filippi of the Engineering Department (UCBM) for specific scientific support provided during the finalization of the paper. Special thanks to Domenico Grigo (former colleague of Eni S.p.A.), who was a master of hydrodynamic modelling and handled fluid pressure studies in sedimentary basins around the world.